\documentstyle[12pt]{article}
\begin{document}
\setlength{\unitlength}{1mm}

{\hfill Preprint DSF-32/94}\\

{\hfill July 1994}\\

\begin{center}
{\Large\bf  Black-Hole Thermodynamics and Renormalization}\\
\end{center}

\bigskip\bigskip

\begin{center}
{\bf Dmitri V. Fursaev}\footnote{e-mail: fursaev@theor.jinrc.dubna.su}
\end{center}

\begin{center}
{{\it Laboratory of Theoretical Physics, Joint Institute for
Nuclear Research, \\
Head Post Office, P.O.Box 79, Moscow, Russia\\
and\\
Dipartimento di Scienze Fisiche, Universit\`a di Napoli -
Federico II -, and INFN\\
Sezione di Napoli, Mostra D'Oltremare Pad. 19, 80125, Napoli, Italy}\\}
\end{center}

\bigskip\bigskip\bigskip

\begin{abstract}
Ultraviolet regime in quantum theory with horizons, contrary to ordinary
theory, depends on the temperature of the system
due to additional surface divergences in the effective action.
We evaluate their general one-loop structure
paying attention to effects of the curvature
of the space-time near the horizon. In particular,
apart from the area term, the entropy of a black hole is shown to
acquire a topological correction in the form of the integral
curvature of the horizon.
To get the entropy, heat capacity and other thermodynamical quantities
finite, such a kind of singularities should be
removed by renormalization of a number of constants in a surface
functional introduced in the effective action at arbitrary
temperature. We also discuss a discrepancy in the different
regularization techniques.
\end{abstract}

\vspace{7cm}

\newpage
\baselineskip=.8cm

\section{Introduction}

Recently, starting from the t' Hooft's paper \cite{a1},
much attention has been paid to one-loop computations
of the entropy in quantum theory on the black-hole space-times
(see \cite{a2}-\cite{a3} and references therein). Different methods employed
on Rindler and Schwarzschild
spaces have indicated a divergent surface term interpreted as
quantum correction to the Bekenstein-Hawking entropy.
Beside this, another term, not reducible to the horizon
area, was pointed out in \cite{a3} for Schwarzschild hole.
However, although a discussion of the particular cases is going on,
the general geometrical form of the additional surface divergences,
their renormalization and related physical aspects are not
completely clear.

A formal reason for surface divergences on a black-hole
space-time can be seen in the following way.
The horizon surface is a set of fixed
points with respect to the one-parameter isometry group associated
with time translations. As a consequence, the corresponding
finite-temperature quantum theory is formulated here in
Euclidean space with conical defects near the horizon,
which results to the surface
divergences in the effective action
similar to those in quantum theory with boundaries \cite{b6}, \cite{b5}.

An interesting feature of new divergences is in their dependence
on temperature. This unusual effect is absent in
ordinary finite-temperature theory where ultraviolet properties
are determined by local geometry and are
not sensible to choice of the quantum state \cite{a16}.
The described features seem to have a universal character for
the bifurcate Killing horizons, including the case of the
cosmological ones \cite{b7}.
A practical interest, however, is the Hawking temperature
when the surface divergences of the effective action or free energy
vanish identically.
Remarkably, it does not exclude
the divergent terms from their derivatives in temperature
and in particular from the entropy.

There is no a unique point of view about the natural cut-off for
such corrections. For instance, one opportunity is to take into
account quantum fluctuations of the horizon \cite{a2}, the
other one is to use the superstring theory \cite{SU}.
Thus,
not excluding these possibilities, it seems to be useful to investigate how
to remove the surface singularities in a way common for quantum field
theory.

The aim of this letter is to represent a general structure of the surface
terms for static spaces with the horizons and to describe their
renormalization. It will be done on the base of the heat kernel expansion
near conical defects \cite{a4}, \cite{a5}. This subject is considered in
sections 2 and 3.
Beside this, we evaluate a
general form of the surface corrections to the black-hole entropy.
Then we conclude with remarks on applicability of the different regularization
schemes.

\section{Divergences}

We are interested in static space-times where thermal
equilibrium can be well defined. In this case the free energy $F(\beta)$
and the one-loop effective action $W(\beta)$ in a scalar theory,
including non-minimal coupling with the curvature, can be written as
\cite{a16}
\begin{equation}
\beta F(\beta)=W(\beta)=\frac 12\log\det(-\Box + \xi R + m^2)=
-\frac 12\int_{0}^{\infty}{ds \over s}Tr K_{M_{\beta}}(s) e^{-m^2s}
\label{eq:log}
\end{equation}
where $\Box$ and $K_{M_{\beta}}(s)$ are the Laplace and the
heat operators on a background manifold
$M_{\beta}$ being a Euclidean section of the corresponding static space.
$\beta$ is the inverse temperature of the system and is the period of the
Euclidean time $\tau$. With respect to the isometry group, associated with
rotations in $\tau$, the 2-dimensional surface $\Sigma$ of the horizon is
a set of the fixed points.

The Hawking temperature $\beta_H^{-1}$ is determined
by the surface gravity $k$ as $\beta_H^{-1}=k/(2\pi)$.
For the sake
of simplicity we will assume that $\beta_H=2\pi$ ($k=1$), then
its right value can be easily restored by changing $\beta$ to
$2\pi\beta/\beta_H$ in all the formulas.
At the Hawking temperature $M_{\beta_H}$ is a smooth manifold.
However, if $\beta \neq \beta _H$, there is a conical singularity at $\Sigma$,
although outside this surface, no matter how close, the geometry
of $M_{\beta}$ is the same as that of $M_{\beta _H}$.
Hence, in comparing to $M_{\beta _H}$ the scalar curvature
acquires a delta-function contribution from $\Sigma$.
Using for it the notation $\bar{R}$ one can write
\begin{equation}
\int_{M_{\beta}}\bar{R}=2(2\pi-\beta)\int_{\Sigma}
+\int_{M_{\beta}} R
\label{cr}
\end{equation}
where $R$ is the local curvature defined by the Riemann tensor in the smooth
region and
the additional surface term is provided by the
conical singularity with deficit angle $2\pi-\beta$
\cite{a3},\cite{a6},\cite{a7}. Note, that for non-minimal coupling
in (\ref{eq:log}) we imply the regular curvature $R$.

As was pointed out in \cite{a7}, it is the integral (\ref{cr}) that should
be used in the Euclidean gravitational action in quantum gravity.
In particular the black hole is an extremum of (\ref{cr})
with subtracted boundary terms.  Besides, the variations of the metric on
$\Sigma$ result to the condition $\beta=2\pi$ corresponding to the smooth
geometry.

The integral in the DeWitt-Schwinger representation (\ref{eq:log}) is
known to diverge on the lower integration limit
as $s\rightarrow 0$. The structure of this divergence can be immediately found
from the asymptotic expansion of $TrK_{M_{\beta}}(s)$, which
reads \cite{a4},\cite{a5}
\begin{equation}
Tr~K_{M_{\beta}}(s)|_{s\rightarrow 0}=
{1 \over (4\pi s)^{d/2}} \sum_{n=0}^{\infty}\left(a_n + a_{\beta,n} \right)
s^n~~~.
\label{eq:singex}
\end{equation}
Here the standard heat coefficients $a_n$, $n\geq1$, given by the integrals on
the powers of the Riemann tensor and its derivatives,
\begin{equation}
a_1=\left(\frac 16 -\xi\right)\int_{M_{\beta}}R~~~,
\label{a1}
\end{equation}
\begin{equation}
a_2=\int_{M_{\beta}}\left({1 \over 180} R_{\mu\nu\lambda\rho}
R^{\mu\nu\lambda\rho}-{1 \over 180}R_{\mu\nu}R^{\mu\nu}
-\frac 16(\frac 15-\xi)\Box R+\frac 12(\frac 16 -\xi)^2R^2\right)~~~, ... ,
\label{a2}
\end{equation}
get modified by the surface terms $a_{\beta,n}$ due to the conical
singularities
near $\Sigma$
\begin{equation}
a_{\beta,1}=\beta c_1(\beta)\int_{\Sigma}~~~,
\label{eq:coeff''}
\end{equation}
$$
a_{\beta,2}=
\beta \int_{\Sigma}\left[c_1(\beta)
\left((\frac 16 -\xi)R+ \frac 16
R_{\mu\nu}n^{\mu}_i n^{\nu}_i -
\frac 13 R_{\mu\nu\lambda\rho}n^{\mu}_in^{\lambda}_in^{\nu}_jn^{\rho}_j
\right)+\right.
$$
\begin{equation}
\left.c_2(\beta)\left(\frac 12
R_{\mu\nu\lambda\rho}n^{\mu}_in^{\lambda}_in^{\nu}_jn^{\rho}_j-
\frac 14 R_{\mu\nu}n^{\mu}_i n^{\nu}_i \right)\right]~~~.
\label{eq:coeff}
\end{equation}
The components of the curvature tensor in (\ref{eq:coeff})
are taken on the surface $\Sigma$ and
$n_i$ are two vectors orthogonal to $\Sigma$ and normalized as
$n^{\mu}_in_{j\mu}=\delta _{ij}$.
$c_{1,2}(\beta)$ in (\ref{eq:coeff''}), (\ref{eq:coeff}) are defined
as
\begin{equation}
c_1(\beta)=\frac 16\left(\left({2\pi \over \beta}\right)^2-1\right)~~~,
{}~~~c_2(\beta)={1 \over 15}c_1(\beta)\left(\left({2\pi \over \beta}\right)^2
+11\right)~~~.
\label{ccoef}
\end{equation}
All other surface coefficients $a_{\beta ,n}$ have the structure similar
to that of (\ref{eq:coeff''}), (\ref{eq:coeff}).

The only coefficient to survive on the Rindler space is
$a_{\beta,1}$ which is proportional to the area of the horizon. On
the other hand,
the next coefficient given by (\ref{eq:coeff}) depends on the curvature
of $M_{\beta}$ near $\Sigma$.
In fact, $a_{\beta,2}$ is defined by three independent geometrical
quantities: by orthogonal components of the Riemann and Ricci tensors,
$R_{\mu\nu\lambda\rho}n^{\mu}_i n^{\lambda}_in^{\nu}_jn^{\rho}_j$,
$R_{\mu\nu}n^{\mu}_in^{\nu}_i$, and by the curvature
$R_{\Sigma}$ of the horizon surface. The latter enters in
the scalar $R$ in (\ref{eq:coeff}), which can be seen from
the Gauss-Codacci equations \cite{schouten}
\begin{equation}
R=R_{\Sigma}+2R_{\mu\nu}n^{\mu}_in^{\nu}_i-
R_{\mu\nu\lambda\rho}n^{\mu}_in^{\lambda}_in^{\nu}_jn^{\rho}_j
-(\chi _{i\mu}^{\mu})^2+\chi _i^{\mu\nu}\chi _{i\mu\nu}
\label{eq:gauss}
\end{equation}
where $\chi _i^{\mu\nu}$ are two second fundamental forms of $\Sigma$
vanishing in the present case due to the isometry.
Thus, for 4-dimensional space $a_{\beta, 2}$ includes a pure topological
term, the integral curvature of $\Sigma$ being the Euler number.
Moreover, as it follows from (\ref{eq:gauss}), for Schwarzschild black
hole and, more generally, for Ricci flat 4-geometries,
$R_{\mu\nu\lambda\rho}n^{\mu}_i n^{\lambda}_in^{\nu}_jn^{\rho}_j=R_{\Sigma}$
and there will be only this topological term in $a_{\beta,2}$.

It is also interesting to find the expressions of these
geometrical quantities
for the Reissner-Nordstrom black hole with
the mass $M$ and charge $Q<M$
\begin{equation}
R_{\mu\nu}n^{\mu}_in^{\nu}_i=-{2r_{-} \over r_{+}^3}~~~,~~~
R_{\mu\nu\lambda\rho}n^{\mu}_in^{\lambda}_in^{\nu}_jn^{\rho}_j=
{2r_{+}-4r_{-} \over r_{+}^3}~~~
\label{rn}
\end{equation}
where, for the unit gravitational constant $G\equiv 1$,
\begin{equation}
r_{\pm}=M\pm(M^2-Q^2)^{1/2}
\label{radius}
\end{equation}
are the radii of the outer and inner horizons
and expressions (\ref{rn}) are evaluated at $r_{+}$. So far as the
horizon area is $4\pi r_{+}^2$ the surface coefficient $a_{\beta,2}$
includes the dimensionless terms proportional to the ratio $r_{-}/r_{+}$.
Note also, that the heat kernel expansion in the form (\ref{eq:singex})
is applicable only when $Q<M$. However it fails for an extremal
hole with $Q=M$ when singularity at $\Sigma$ is not more conical. In this case
the Hawking temperature $\beta_{H}^{-1}$ is zero and polynomial coefficients
$c_{1,2}(\beta)$, which really depend on dimensionless combinations
$(\beta_H/\beta)^2$, turn out to be infinite at $\beta\neq\beta_H$.

The divergent part $W_{div}(\beta)$ of the effective action (\ref{eq:log})
can be easily obtained
from (\ref{eq:singex}). For example, in dimensional
regularization \cite{a16} one can
represent $W_{div}(\beta)$
as a sum of the volume, $W_{div,~vol}$, and surface, $W_{div,~surf}$, parts
$$
W_{div}(\beta)= W_{div,~vol}(\beta)+W_{div,~surf}(\beta)~~~,
$$
\begin{equation}
W_{div,~vol}(\beta)=-{1 \over 32\pi ^2\epsilon}
\left[{(m^2)^2 \over 2}a_0-m^2 a_1
+a_2\right]~~~,
\label{vol}
\end{equation}
\begin{equation}
W_{div,~surf}(\beta)=-{1 \over 32\pi ^2\epsilon}
\left[-m^2 a_{\beta,1} +a_{\beta, 2}\right]~~~
\label{surf}
\end{equation}
where $\epsilon$ is a regularization parameter interpreted as extra dimensions
of the space.
As it follows from (\ref{vol}), the volume integral
$W_{div,~vol}(\beta)$ represents
standard ultraviolet divergences in quantum theory in curved space-time.
Its dependence on temperature is trivial; $\beta$ enters in
$W_{div,~vol}(\beta)$
as a multiplier. Hence, $W_{div,~vol}(\beta)$ results to an infinite
correction to the vacuum energy of the system,
but does not contribute to its entropy.
The properties of the functional $W_{div,~surf}(\beta)$ are
different. It is given on the horizon $\Sigma$ and
describes the surface divergences induced by conical singularities of
$M_{\beta}$. Moreover, the equations
show that in the free energy
new divergences are the forth order polynomials in temperature and therefore
they do contribute to the entropy. This interesting feature of quantum
fields in presence of the horizons is absent in standard quantum
theory.

\section{Renormalization}

In black-hole thermodynamics one is interested in the Hawking temperature
$(\beta=2\pi)$
when $M_{\beta}$ is a smooth manifold and all the surface terms vanish
in the effective action. However,
as follows from (\ref{eq:coeff''})-(\ref{ccoef}) and (\ref{surf}),
the derivatives in $\beta$ of $W_{surf,~div}(\beta)$ are not equal to
zero at $\beta=2\pi$. Much attention has been paid to the fact that
it results to the infinite corrections to the entropy of a black hole.
For Rindler space-time these infinities can be removed from the entropy
by renormalization of the gravitational constant \cite{SU}.
For the Schwarzschild geometry the additional divergence due to the
curvature can be eliminated by renormalization of the other coupling
constant in the one-loop gravitational action \cite{a3}.

These prescriptions do not seem to have a universal character so far as
they only concern the entropy, but leave the divergences in the other
physical quantities related with the higher derivatives in $\beta$
of the effective action.
For instance, the heat capacity depends on the second derivative
and it can be shown to acquire the surface term at the Hawking temperature
different from that of the entropy.
It means that for complete solution of the renormalization
problem one should renormalize the effective action at an {\it arbitrary}
parameter $\beta$, which could provide then finite thermodynamical
quantities.

It is obvious that surface divergences cannot be removed by renormalization
of some constants in the classical action because the latter does not have
the surface terms. Such terms could appear in result of a specific matter
distribution
over the horizon $\Sigma$, which is not the case of classical black holes.
Consequently, to remove such a kind of singularities one should introduce
in the effective action a surface functional vanishing at $\beta=2\pi$.
In scalar theory its most general structure should be
$$
W_{surf}(\beta)=b_0(2\pi-\beta)\int_{\Sigma} +
\beta \int_{\Sigma}\left(b_1c_1(\beta)+b_2c_1(\beta) R_{\Sigma}+\right.
$$
\begin{equation}
\left. (b_3 c_1(\beta) +b_4c_2(\beta)) R_{\mu\nu}n^{\mu}_in^{\nu}_i+
(b_5 c_1(\beta) +b_6c_2(\beta))
R_{\mu\nu\lambda\rho}n^{\mu}_in^{\lambda}_in^{\nu}_jn^{\rho}_j\right)~~~
\label{surfact}
\end{equation}
where $b_i$ are undetermined constants that are not depend on the background
geometry.

A remark about the coefficient $b_0$ is in order. The reason for
this term is in renormalization of the gravitational constant $G$.
Indeed, the classical Einstein action should be determined by the integral
curvature (\ref{cr}) and it includes an additional contribution
from the conical singularity. On the other
hand, due to the heat coefficient $a_1$ the divergent term in (\ref{vol})
depends only on the local scalar curvature $R$. Thus this divergence
is removed by the renormalization both $G$ and $b_0$ as follows
$$
-{1 \over 16\pi G}\int_{M_{\beta}}\bar{R}+
b_0(2\pi-\beta) \int_{\Sigma}+~ {m^2 (1/6-\xi)\over 32\pi ^2 \epsilon}
\int_{M_{\beta}}R=
$$
\begin{equation}
=-{1 \over 16\pi G_{ren}}\int_{M_{\beta}}\bar{R}+b_{0,ren}(2\pi-\beta)\int_
{\Sigma}~~~.
\label{ren}
\end{equation}
Besides, in the first order of the Planck constant
the bare parameters $G$, $b_0$ are expressed through the renormalized
ones $G_{ren}$, $b_{0,ren}$ as
\begin{equation}
G=G_{ren}-{1 \over \epsilon}(\frac 16 -\xi){m^2 \over 2\pi}G^2_{ren}~~~,
\label{ren1}
\end{equation}
\begin{equation}
b_0=b_{0,ren}+{m^2 (1/6-\xi) \over 16\pi ^2 \epsilon}
\label{ref2}
\end{equation}
and renormalization of $G$ is not influenced by the surface divergences.

To get rid off the surface divergences, $W_{div,~surf}$, eq. (\ref{surf}),
other constants $b_i$ in $W_{surf}$ are renormalized similar to $b_0$ in
such a way that renormalization equations do not depend on the background
metric. The number of these constants corresponds to a freedom in the
renormalization recipe. Indeed, the
surface action must retain isometry invariance and
reparametrization invariance of coordinates on $\Sigma$. Hence,
in a 4-dimensional theory it can be characterized, apart from a
trivial constant,
by three invariants $R_{\Sigma}$, $R_{\mu\nu}n^{\mu}_in^{\nu}_i$ and
$R_{\mu\nu\lambda\rho}n^{\mu}_in^{\lambda}_in^{\nu}_jn^{\rho}_j$.
On the other hand,
one should remember that origin of $W_{div,surf}$ is in specific
geometry of $M_{\beta}$ near $\Sigma$, which enables to
find out a form of its dependence on $\beta$.
After the scale transformation of the
Euclidean time $\tau=\beta\tau '$ the parameter $\beta$
appears only in the time component of the metric tensor $g_{\tau '
\tau '}=g_{\tau\tau}\beta ^{-2}$ and, consequently, the surface terms
should have the structure $\beta P(\beta ^{-2})$ where $P(\beta ^{-2})$
is a polynomial vanishing at $\beta=2\pi$ (additional factor $\beta$
comes from the volume element in the action integral). As follows
from (\ref{surf}), $P(\beta^{-2})$ is a second order
polynomial that can be decomposed in $c_1(\beta)$ and
$c_2(\beta)$.
These arguments and the expressions of
$a_{\beta,1}$,
$a_{\beta,2}$ (see (\ref{eq:coeff''}), (\ref{eq:coeff})) define the
functional $W_{surf}(\beta)$ in the form (\ref{surfact}).

After renormalization one obtains a finite but undetermined contribution
to the entropy $S_{surf}(\beta)$ due to $W_{surf}(\beta)$
$$
S_{surf}(\beta)=
(\beta {\partial \over \partial \beta} -1)W_{surf}(\beta)=
$$
\begin{equation}
=-2\pi b_0\int_{\Sigma}+
\beta ^{-1}\int_{\Sigma}\left[\gamma _1 + \gamma _2 R_{\Sigma}+
(\gamma _3  +  \gamma _4 \beta ^{-2}) R_{\mu\nu}n^{\mu}_in^{\nu}_i+
(\gamma _5 + \gamma _6 \beta ^{-2})
R_{\mu\nu\lambda\rho}n^{\mu}_in^{\lambda}_in^{\nu}_jn^{\rho}_j\right]~~~
\label{entrcor}
\end{equation}
where $\gamma _i$ are some numerical combinations from renormalized
constants $b_i$.
Thus at the Hawking temperature $S_{surf}(\beta=2\pi)$ is characterized
by four geometrical terms and four corresponding constants.
One term, determined by the horizon area $A=\int_{\Sigma}$,
was found and discussed in the literature for some particular cases
\cite{a1}-\cite{b4}. The functional (\ref{entrcor}) indicates also other
corrections not reducible to the area and depending on the
curvature near the horizon. For instance, for Schwarzschild black hole there is
a topological term $\int_{\Sigma}R_{\Sigma}$ in the entropy (cf. \cite{a3}),
and for the charged holes there are other terms that can be expressed through
the mass $M$ and charge $Q$ with the help of (\ref{rn}), (\ref{radius}).
Note also, that in the vacuum state ($\beta\rightarrow\infty$) only
the first term in (\ref{entrcor}) contributes to the entropy.

Finally, one can proceed in this way and use (\ref{entrcor}) to calculate the
surface corrections to the heat capacity of the system, $C_{surf}(\beta)=
-\beta {\partial \over \partial\beta}S_{surf}(\beta)$, and to other
thermodynamical quantities.

\section{Conclusions and remarks}

We presented a general form of the surface divergences in the effective action
and entropy on static spaces with the horizons. Our conclusion is that
this kind of divergences is not reduced only to the area of the horizon
but has a more complicated structure depending on the curvature of the
space-time. We point out that renormalization of the gravitational constant
and other couplings in the one-loop gravitational action is not sufficient to
remove these divergences from the all thermodynamical quantities.
To this aim one has to introduce a surface functional (\ref{surfact})
with a number of constants that might be fixed in a more fundamental theory.

The fact that the entropy corrections (\ref{entrcor}) have a more rich form
then the Bekenstein-Hawking entropy is not surprising. It should be
borne in mind that the one-loop action in quantum gravity includes
the second order terms in the curvature, like those in (\ref{a2}),
which must be taken into account in evaluation of the entropy.
{}From this point of view it would be interesting to analyze the relation
between the coupling constants $b_0,~\gamma_i$ in (\ref{entrcor})
and those in the one-loop gravitational action. A relevant approach
can be found in \cite{a3}.

A remark is in order about different techniques to be used to renormalize
the surface divergences. We employed here the heat-kernel method.
Other approach is to assume that the wave functions vanish within some
fixed distance from the horizon. In this case all the integrals stop short
before the horizon and turn out to be finite. Such method was initially
proposed in \cite{a1} as the "brick-wall" model and it is used
with some modifications in other papers.

The heat-kernel and the brick-wall methods result to the similar structure
of the surface divergences, but they are not quite equivalent. An analysis
shows that they give different dependence on temperature of the
of the surface terms in free energy.
The reason for this is in their topological properties \cite{a18}.
The heat-kernel approach does not change the topology of the cone.
On the other hand, the brick-wall method is equivalent to cutting of the cone
tip, which changes the topology to $S^1\times R^{+}$.
Thus, from this point of view the heat-kernel technique presented here
seems to be more preferable, although for computation of the local renormalized
quantities outside the horizon one can use both them.

In the third method that should be mentioned the cone tip is changed to
a smooth manifold \cite{a3}. It holds the topology and does not contradict
to the heat-kernel results. However, the surface divergences must be
extracted from the non-local part of the effective action, which
complicates the computations.

Note finally, that our consideration was confined to the static spaces.
An interesting problem is to analyze the same aspects for
the stationary black-hole geometries.

\bigskip\bigskip

I would like to thank Sergey Solodukhin for valuable remarks and
Franco Buccella and Gennaro Miele for warm
hospitality during my visit in the University of Naples.
This work was supported in part by the grant RFL000 of
the International Science Foundation.

\end{document}